\documentclass{aastex63}
\usepackage{epstopdf}
\newcommand{\teff}{$T_{\textrm{eff}}$}

\shorttitle{new moving group}
\shortauthors{Liang et al.}

\begin{document}
\title{A new moving group in the Local Arm}

\correspondingauthor{Jingkun Zhao}
\email{zjk@nao.cas.cn}

\author[0000-0001-9283-8334]{Xilong Liang}
\altaffiliation{Key Laboratory of Optical Astronomy, National Astronomical Observatories, Chinese Academy of Sciences, Beijing 100012, China}
\altaffiliation{School of Astronomy and Space Science, University of Chinese Academy of Sciences, Beijing 100049, China}
\author[0000-0003-2868-8276]{Jingkun Zhao}
\altaffiliation{Key Laboratory of Optical Astronomy, National Astronomical Observatories, Chinese Academy of Sciences, Beijing 100012, China}
\altaffiliation{School of Astronomy and Space Science, University of Chinese Academy of Sciences, Beijing 100049, China}
\author[0000-0002-8442-901X]{Yuqin Chen}
\altaffiliation{Key Laboratory of Optical Astronomy, National Astronomical Observatories, Chinese Academy of Sciences, Beijing 100012, China}
\altaffiliation{School of Astronomy and Space Science, University of Chinese Academy of Sciences, Beijing 100049, China}
\author{Xiangshong Fang}
\altaffiliation{Key Laboratory of Optical Astronomy, National Astronomical Observatories, Chinese Academy of Sciences, Beijing 100012, China}
\author{Xianhao Ye}
\altaffiliation{Key Laboratory of Optical Astronomy, National Astronomical Observatories, Chinese Academy of Sciences, Beijing 100012, China}
\altaffiliation{School of Astronomy and Space Science, University of Chinese Academy of Sciences, Beijing 100049, China}
\author{Jiajun Zhang}
\altaffiliation{Key Laboratory of Optical Astronomy, National Astronomical Observatories, Chinese Academy of Sciences, Beijing 100012, China}
\altaffiliation{School of Astronomy and Space Science, University of Chinese Academy of Sciences, Beijing 100049, China}
\author{Xiaoming Kong}
\altaffiliation{School of Mechanical, Electrical and Information Engineering, Shangdong University at Weihai, Weihai 264209, China}
\author[0000-0002-8980-945X]{Gang Zhao}
\altaffiliation{Key Laboratory of Optical Astronomy, National Astronomical Observatories, Chinese Academy of Sciences, Beijing 100012, China}
\altaffiliation{School of Astronomy and Space Science, University of Chinese Academy of Sciences, Beijing 100049, China}

\begin{abstract}

We present a new moving group clustered in kinematics, spatial position and elemental abundances. Its spatial position is around the center of the Local Arm of the Milky Way. A convergent point method was taken to select candidate member stars.\textbf{ Among 206 candidate member stars, 74 are pre-main-sequence stars and some of them have stellar disks.} We presume those pre-main sequence stars belong to Orion nebula. We suggest this moving group is caused by density wave of the Local Arm passing by.

\end{abstract}
\keywords{Galaxy kinematics, Stellar associations, Galaxy abundances}

\section{Introduction}

The existence of moving groups in the solar neighbourhood has been known for over one century and they are believed from disrupted stellar clusters \citep{kap05,egg58,sku97} at the begining. Many moving groups got their names by their convergent points from a series papers of \citet{egg58}. However, Dehnen (1998) pointed out that most moving groups in the solar neighborhood could be caused by orbital resonances. Later observations indicate dynamical effects of non-symmetric parts of the Galaxy may be responsible for most of the best known structures, such as the Sirius, Hyades, Pleiades, Hercules and $\gamma$ Leo moving group \citep{fam05,fam07,ben07,fam08,lia18}. With Gaia DR2 data release, many new kinematic sub-structures in the solar neighbourhood were found \citep{kat18,ant18,ram18} which have been thought to be related to phase mixing. Nonetheless, there are still some moving groups belonging to dispersed stellar clusters. \citet{sil07} concluded that HR 1614 moving group is the remnant of a dispersed star-forming event according to its homogenous age and abundances. At least a part of Sirius moving group is thought to be composed of remnants evaporated from a stellar cluster \citep{kle08,lia17}. Besides those known moving groups \citep{fah18} which might be parts of larger kinematic picture, there are also many new moving groups lately found \citep{jos08,gal13,ros18,gol18,yeh19} close to star clusters or star forming regions. As is known, clusters will disperse in response to many events such as sudden changes in mass driven by supernova or stellar wind, tidal interactions with gravitational field or internal relaxation. As clusters disperse, low mass stars are preferentially lost, as they develop higher velocities than more massive cluster members. Some of the remnants of associations are becoming young moving groups before totally dissolving into the field, such as the Scorpius-Centaurus OB-association has been studied as a moving group over a century. Many sophisticated codes have been developed to identify moving groups and their members comprehensively as possible in the solar neighborhood such as ASYA \citep{tor16}, LACEwING \citep{rie17} and BANYAN \citep{mal14,gag14,gag16,gag18}. However, the identification of star members of moving groups is not easy by only kinematics and photometry. \textbf{It had better include additional chemical information.} Chemical tagging technique \citep{fre02} is a credible way to judge whether a moving group origin from a stellar cluster. Some well known moving groups have been studied with detailed chemical abundances from high resolution spectra \citep{ben07,sil07,bub10,sil11,bia12,tab12,sil13,mon16,tab17,lia18,zhao18}. However, the number of stars that can be observed with high resolution spectroscopy is still small. Therefore we turned to machine learning to obtain elemental abundances from low resolution spectra.

The Orion complex is the nearest site of active star forming region and it contains multiple stellar populations and complex substructures such as clusters, OB associations and young stellar moving groups. Within star forming Orion A and B molecular clouds, there are massive star clusters such as the Orion Nebula Cluster (ONC), NGC 2024, and NGC 2068. The Orion Nebula Cluster is significantly younger than other regions in Orion A, while Orion A is younger than the still-larger Orion D region. Consequently, there is a gradient of increasing mean age and age spread as one moves from denser to less dense regions \citep{get14,get18,zar19}. Star forming regions are hierarchically structured, containing both dense parts for which mass removal is slow compared to the local dynamical time, and diffuse parts for which it is fast \citep{kru19}. \citet{kou18} found the Orion D group is in the process of expanding while Orion B is still in the process of contraction. \citet{kou18} also found the proper motions of $\lambda$ Ori are consistent with a radial expansion due to an explosion from a supernova. \citet{zar19} confirmed multiple events caused kinematic and physical sub-structure rather than a simple sequential scenario. There are runaway stellar groups while there are stars going through gravitational infall \citep{mcb19,get19}. Substructures inside the Orion nebula has been comprehensively studied by former researches and it is not analysed in this paper.

In Section 2, we describe the data and method used to identify a moving group. Section 3 presents candidate member stars of the new foumd moving group in parameter space. In subsection 3.4, we compared the new found moving group with known young stellar moving group OrionX. Finally, the main outcomes of our work and perspectives for the future are summarized in Section 4.

\section{Data and method}

The data we used are 657 561 stars with machine learning chemical abundances of target sample from \citet{lia19}, which are common stars of LAMOST DR5 catalogue \citep{luo15,zhao12} cross-matched with Gaia DR2 \citep{gai18} catalog after removed duplicated sources. They are mainly G, K gaints with \teff  between 4000K and 5500K. With parallax and proper motion from Gaia and radial velocity from LAMOST, three dimensional velocity components for stars were obtained. We calculated position \textit{xyz} and their velocity components \textit{UVW} in the right handed rectangular coordinate system with origin at the Galactic Center and axis parallel to the Local Standard of Rest(LSR). In this coordinate system, \textit{x} axis points from the Sun toward the Galactic center, \textit{y} axis points along the direction of Galactic rotation of LSR, and \textit{z} axis points toward the north Galactic pole. The Sun is placed at height $z=0.014$ kpc \citep{bin97} and galactic radius $R = 8.34$ kpc with the circular rotation speed $ V_{c}=240$ km s$^{-1}$ \citep{rei14}. The peculiar velocities of the Sun with respect to the LSR are taken as (11.1, 12.24, 7.25) km s$^{-1}$ \citep{sch10}. To show over-densities in the velocity space, wavelet transform technique \citep{lia17} is adopted. Since velocity substructures of stars within 200 pc have been studied extensively \citep{fah18}, we choose to look into stars a little father away. We selected a sample of 32 111 stars with $-8.54$ kpc $>  x > -9.04$ kpc, $\mid y \mid < 0.5$ kpc and $\mid z \mid  < 0.5$ kpc.

A convergent point method \citep{bro50,jon71} has been used to select member stars of moving group. The classical convergent point method is based on common space motion of member stars in a moving group \textbf{resulting} in converging proper motions and many moving groups got their names from the convergent point. But our method is a little different from the classical convergent point method. We adopted all three components of velocity converging in a real position rather than two proper motion components converging in a projected point on the celestial sphere. Our method mainly \textbf{includes }two steps. First we calculated the sum of the distance from a given point to the reverse extension line of each star's velocity vector over all stars in the data sample and then minimize the sum to obtain the convergent point. After this, we removed those stars with largest distances which apparently do not belong to the new moving group.\textbf{ The procedure is repeated }till the largest distance \textbf{drops} in an acceptable range. Subsequently, all non-rejected stars are identified as members. Thus this procedure simultaneously determined member stars and the convergent point which might imply the birth place. This method assumes that all member stars are moving away from one point. In fact strict convergence to one point is not necessary because of measurement errors and the internal velocity dispersion of a moving group.

\section{New moving group}
\subsection{Count density in \textit{U, V} coordinate}
\begin{figure}[tpb]
\begin{center}
\plottwo{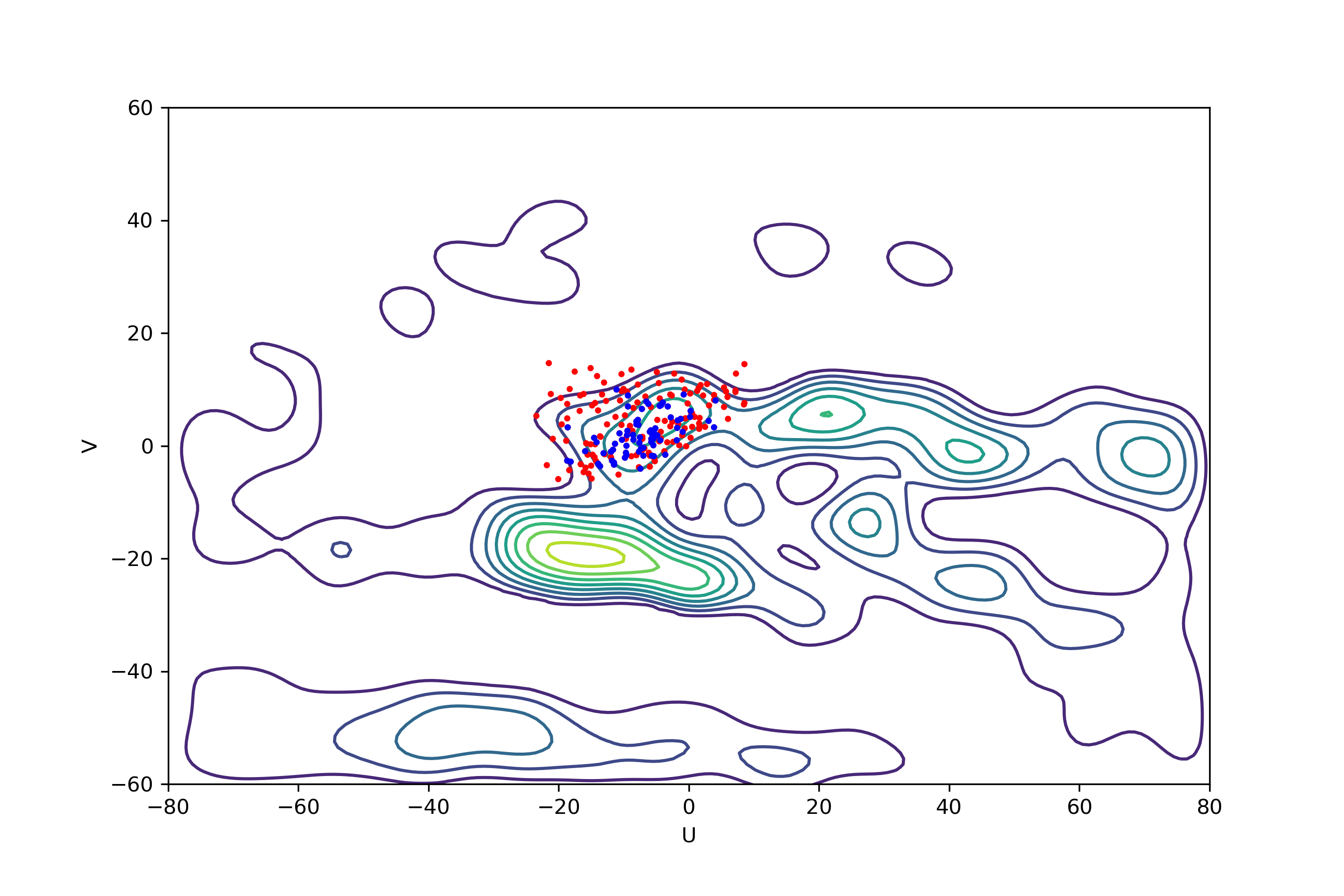}{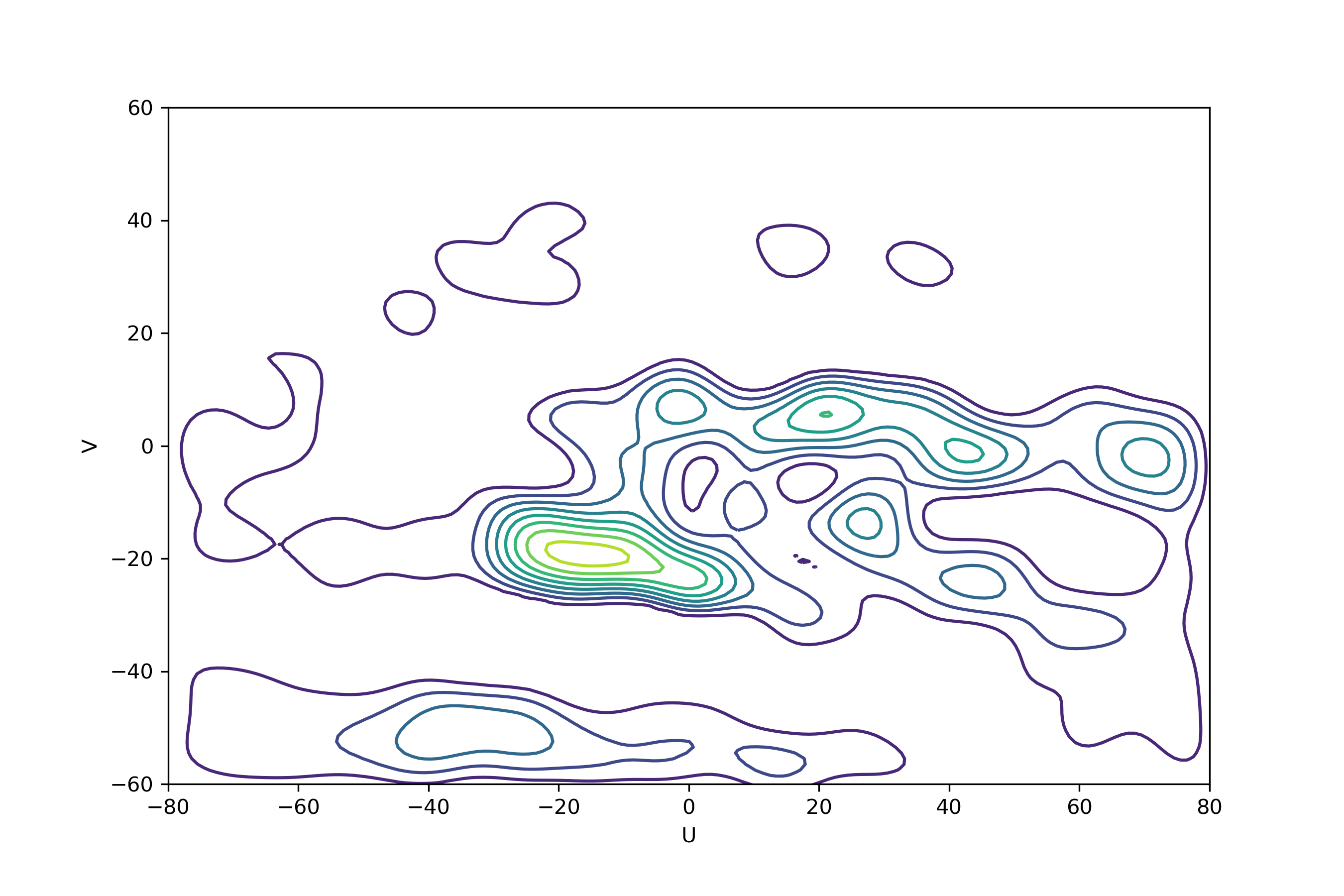}
\end{center}
\caption{Wavelet transform of \textit{UV} velocity distribution of the selected star sample. The left subplot shows sample with pre-main sequence stars, while the right subplot shows sample without pre-main sequence stars. Those points in the left subplot are candidate members selected by convergent point method in which blue points are pre-main sequence stars.\label{fig:uv}}
\end{figure}

In figure \ref{fig:uv}, the left subplot shows contour of wavelet transform of \textit{U, V} density distribution of the selected star sample, while the right subplot shows contour of wavelet transform of \textit{U, V} density distribution of the selected star sample without pre-mains sequence stars (explained later). The median uncertainties values of \textit{U, V} and \textit{W} are respectively 3.7, 3.3 and 2.5 km/s of the selected star sample. Spatial region of this selected sample is similar to region 3 in \citet{lia17} and it contains the spatial region of V6 from \citet{ram18}. Therefore our contours look very similar to subplot (b) of figure 5 of \citet{lia17} and subplot (f) of figure 5 of \citet{ram18}. However, the peak around (0, 7) km/s in our figure is much more significant than in theirs which are only in dense regions but not density peaks. We call it a new moving group candidate since it is so obvious for the first time. The difference between our data and \citet{lia17} lies in\textbf{ that} this sample is selected from LAMOST DR5 cross-matched with Gaia DR2 rather than Gaia DR1 TGAS. The difference between our data and \citet{ram18} mainly lies in \textbf{that} \citet{ram18} used the Gaia DR2 sample and we used only G,K giants with radial velocity from LAMOST. There are about 1868 stars around the peak in the \textit{UV} coordinate of our sample. After applied the convergent point method we got 206 stars as candidate members in which 74 stars are pre-main sequence stars. Velocity directions of those member stars converge within about 100 pc around (x, y, z)$ = (-8.70, -0.15, -0.12)$ kpc. As shown in figure \ref{fig:uv}, those points in the left subplot are selected candidate members in which red dots represent normal giants while blue dots represent pre-main sequence stars. Pre-main sequence stars take up about one third of candidate members, therefore we thought this over-density might be caused by pre-main sequence stars at first. However, as the right subplot in figure \ref{fig:uv} shows, that over-density peak is still there after we removed those pre-main sequence stars. Thus we think pre-main sequence stars only contribute partly to the number density peak in that velocity region.

\subsection{Effective temperature and gravitational acceleration}
\begin{figure}[tpb]
\epsscale{.80}
\plotone{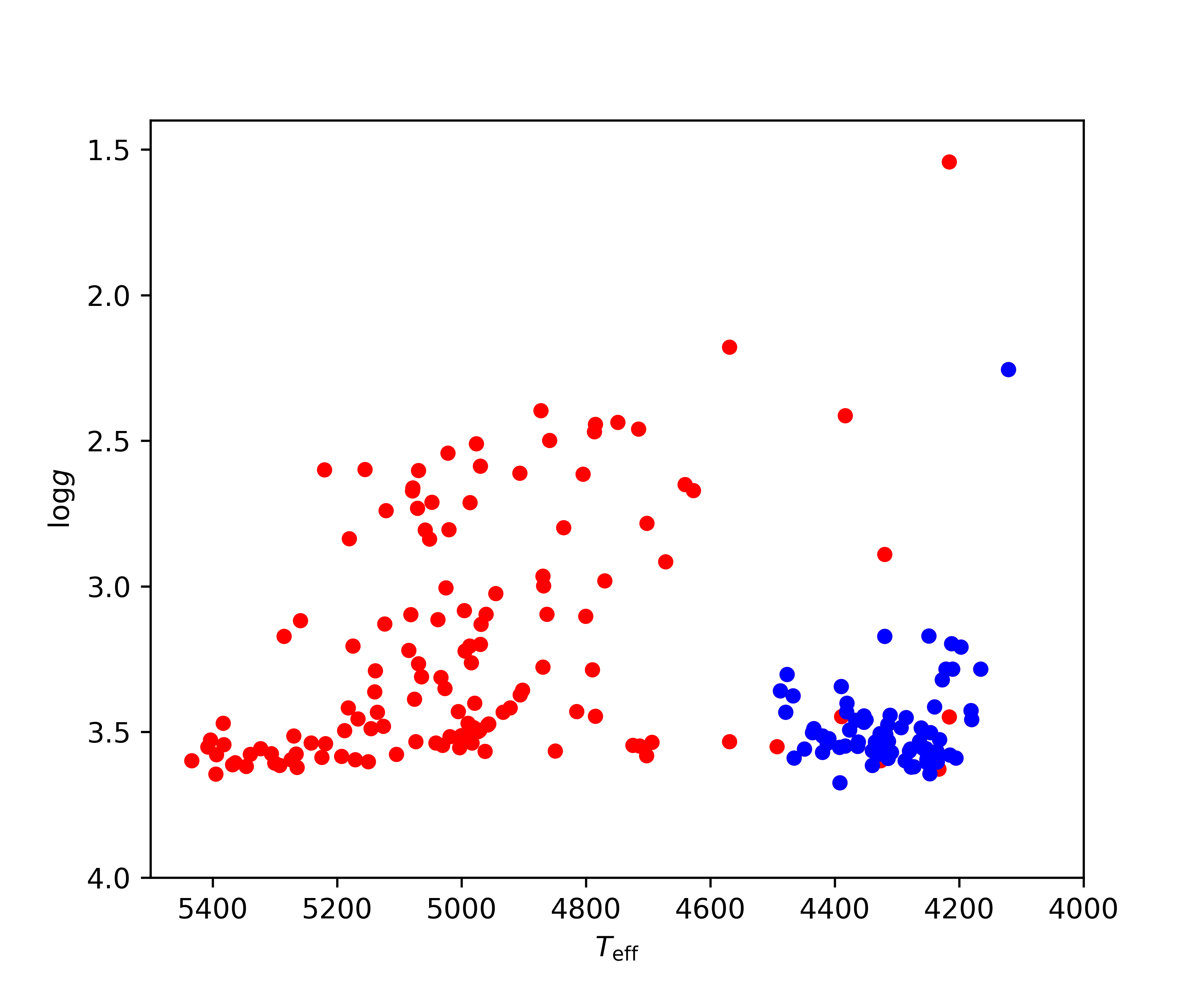}
\caption{$T_{\textrm{eff}}$ versus log$ g$ distribution of candidate members.\label{fig:tg}}
\end{figure}

Figure \ref{fig:tg} shows \teff  versus log $g$ distribution of candidate members. Red dots represent normal giants and blue dots represent pre-main sequence stars for all figures in this paper. To make sure pre-main sequence stars are rightly classified, we checked spectra of those stars at the region of pre-main sequence stars on figure \ref{fig:tg}. Those apparently not pre-main sequence stars were kicked out as normal stars. Spectra of those pre-main sequence stars have hydrogen emission lines and some have forbidden lines of elements like nitrogen, sulfur and so on. Table \ref{tab:liabu} in the Appendix lists our measured equivalent widths of lithium 6707 \AA  ~along with snrr from LAMOST spectra. Since we used low resolution spectra, those equivalent widths are very rough estimations. Those pre-main sequence stars are naturally much younger than G, K giants. We speculate they are clustered together for different reasons.

\subsection{Chemical abundances}

\begin{figure}[tpb]
\epsscale{.80}
\plotone{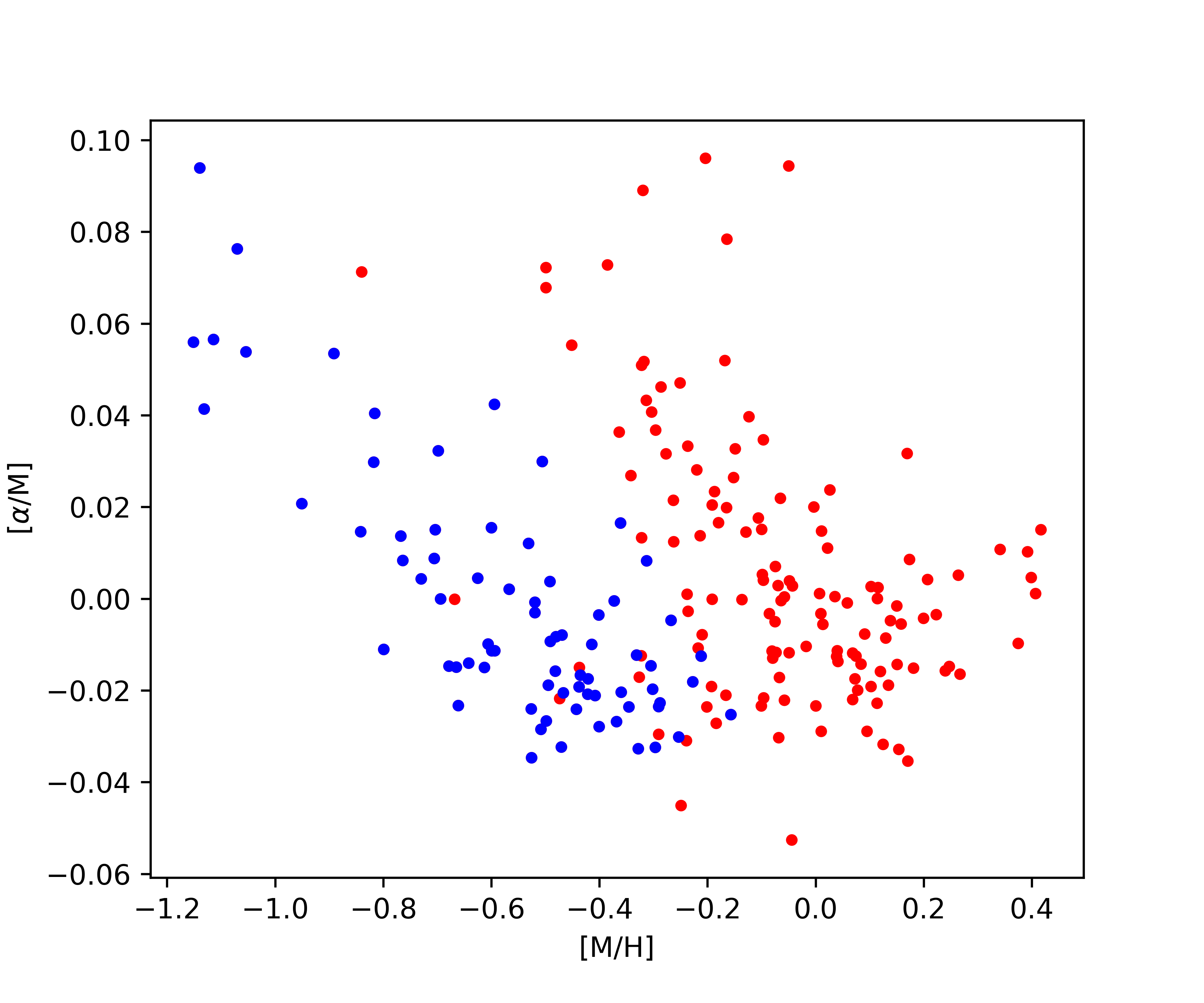}
\caption{[$\alpha$/M] versus [M/H] distribution of candidate members. Red dots are G, K giants while blue points are pre-main sequence stars.\label{fig:amh}}
\end{figure}

Figure \ref{fig:amh} shows [$\alpha$/M] versus [M/H] distribution of candidate members of the new moving group. The median metallicity value of pre-main sequence stars is -0.506 while the median metallicity value of giants is -0.068. There is -0.438 dex difference between them but metallicity values of pre-main sequence stars usually been systematically underestimated. According to spatial positions, those pre-main sequence stars should belong to Orion nebula and their metallicity should be solar abundance. For spectra of many pre-main sequence stars, there are emission lines within absorption lines and some lines become totally emission lines. Since the neural network in \citet{lia19} is dominated by most stars with absorption lines, elemental abundances shall be wrongly estimated for stars with emission lines. Figure \ref{fig:gp} in the appendix shows parts of spectral features of a pre-main sequence star of member stars as an example. Those forbidden lines and wide H$\alpha$ line indicate this star may have a stellar disk.

\begin{figure}[tpb]
\epsscale{}
\plotone{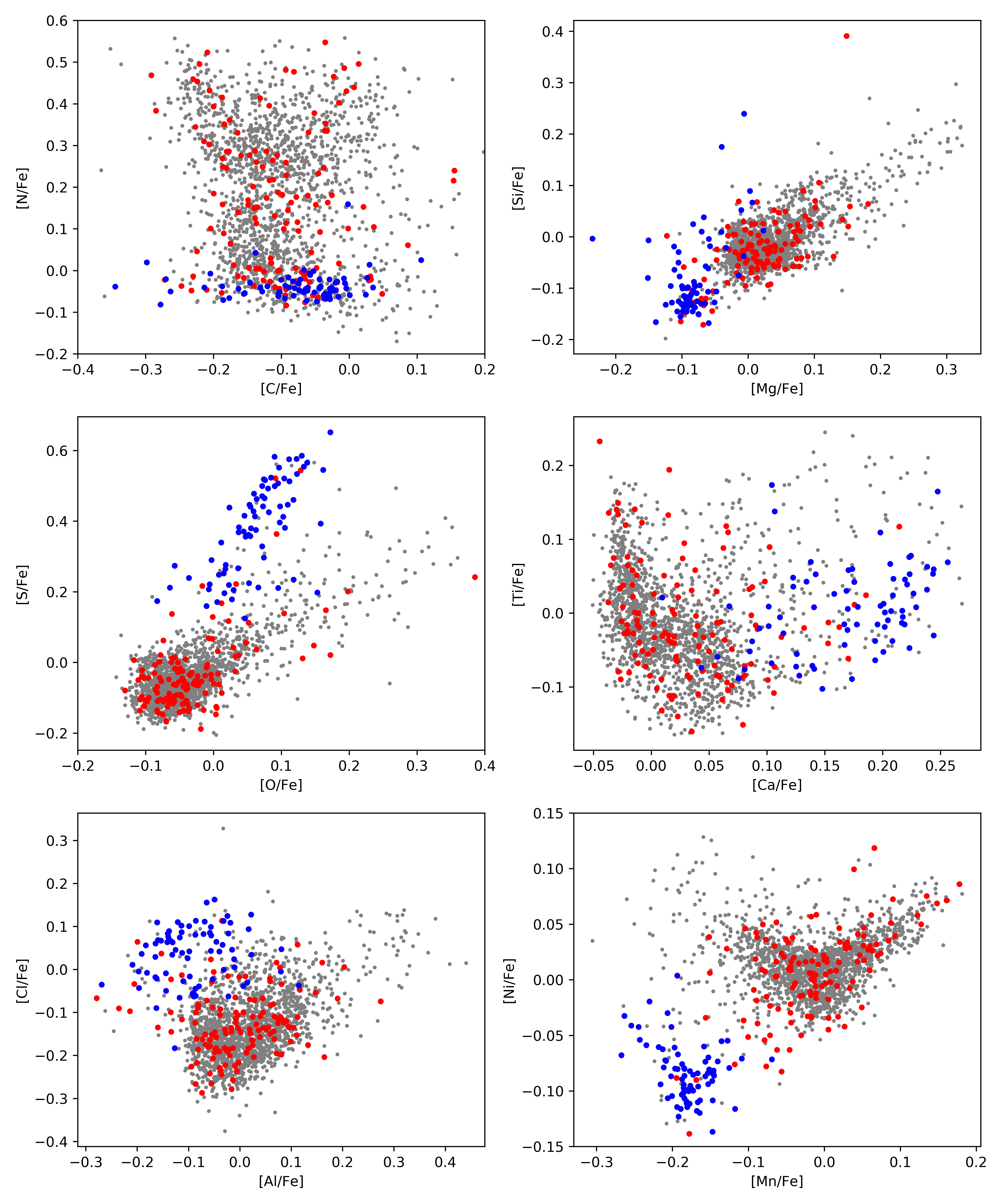}
\caption{Abundances distributions of candidate members. Red dots are G, K giants while blue dots are pre-main sequence stars.\label{fig:abu}}
\end{figure}

In figure \ref{fig:abu}, we show other elemental abundances distributions of candidate members of this new moving group. Red dots are candidate member stars in which blue dots are pre-main sequence stars. While background grey dots are all stars around the density peak of wavelet transform as comparison. It's clear that elemental abundances distributions of candidate members are too scattered to be from a cluster. Chemical distributions of those pre-main sequence stars cluster together although there might be systematic errors for each elemental abundance. Since those pre-main sequence stars should all come from orion nebula, their elemental abundances should be like solar abundances. Each emission lines have different effect on different elements, therefore chemical abundances of pre-main sequence stars have been wrongly estimated in various degrees relative to solar abundances. \textbf{Since the neural network in \citet{lia19} is not specially designed for pre-main sequence stars, so abundances for pre-main sequence stars are not accurate but their elemental abundances do clump together. Thus we suggest machine learning can be used to classify pre-main sequence stars.} For those giants of the moving group, they are not from a stellar cluster. We think they have been clustered before pre-main sequence stars were born by some other reason.

\subsection{Spatial distribution and other velocity distributions}

\begin{figure}[tpb]
\plotone{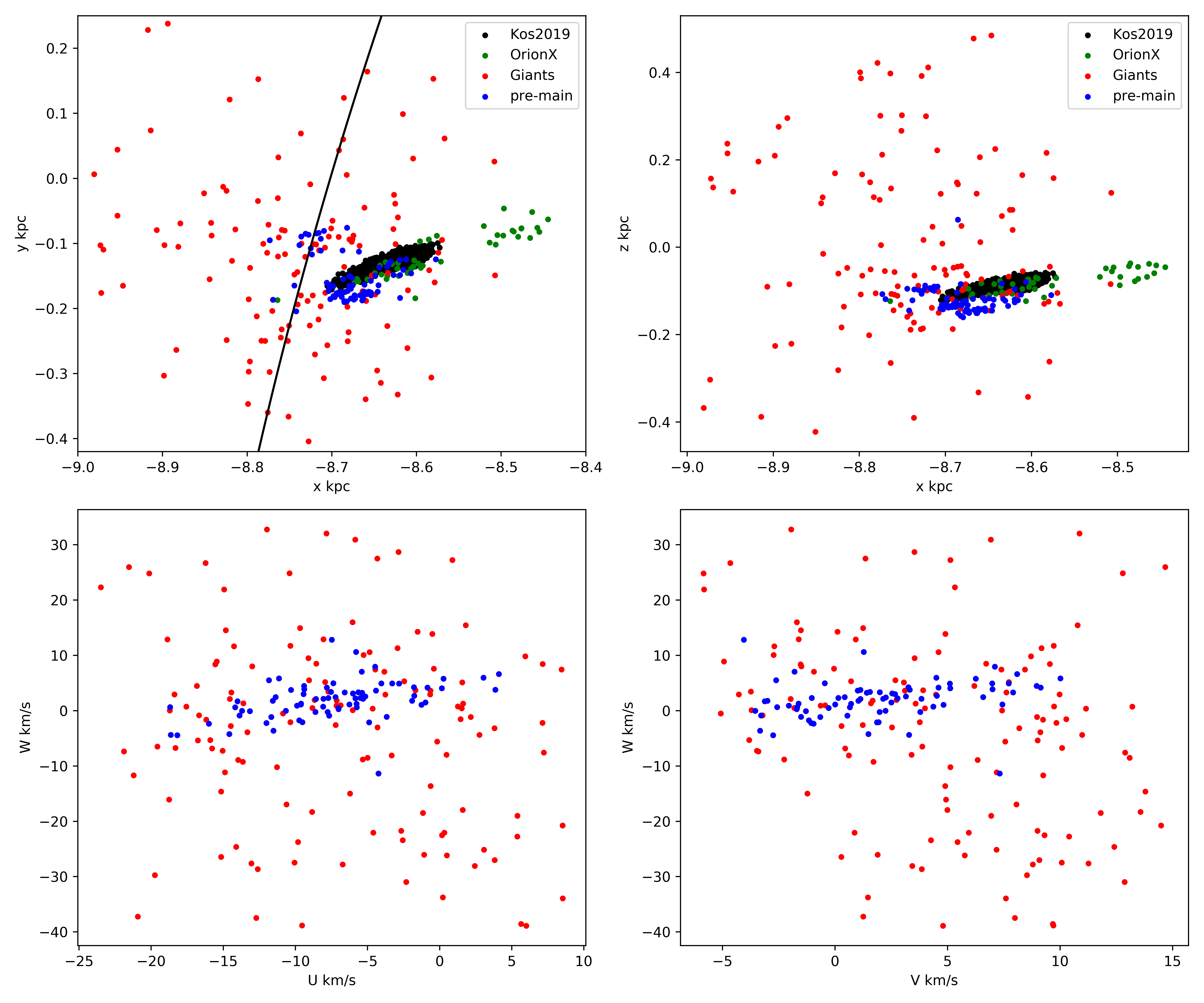}
\caption{Spatial and other velocity distribution of candidate members.\label{fig:yz}}
\end{figure}

Figure \ref{fig:yz} shows distributions of candidate members in spatial coordinates and other velocity coordinates. The top two subplots respectively show \textit{y} versus \textit{x} and \textit{z} versus \textit{x}, while bottom two subplots respectively show \textit{W} versus \textit{U} and \textit{W} versus \textit{V}. \textbf{Green dots represent member stars of OrionX from \citet{bou15} while black dots represent stars listed in the appendix of \citet{kos19} which shows the spatial position of Orion nebula.} It can be seen those blue pre-main sequence stars distributes more compact than red normal giants and they are surrounded by normal giants in both spatial coordinates and velocity coordinates. According to spatial positions, those pre-main sequence stars should belong to the Orion nebula. Pre-main sequence stars in star forming region as the Orion nebula have been extensively studied by former researches. Therefore, this paper focuses on G, K giants of candidate members which have not been studied. The black curve in the \textit{y} versus \textit{x} subplot represents the center of the Local Arm of our Galaxy from \citet{rei14}. As can be seen, those candidate members are around the center of the Local Arm. We think this new moving group should be related to the Local Arm. Those pre-main sequence stars from the new moving group are experiencing diffusion from their born place. While for those G, K giants, it could be density wave that drove them to be there with clumped velocity. Now candidate members of the new moving group are moving away from convergent point.

\subsection{Comparison with OrionX}
\begin{figure}[tpb]
\epsscale{.60}
\plotone{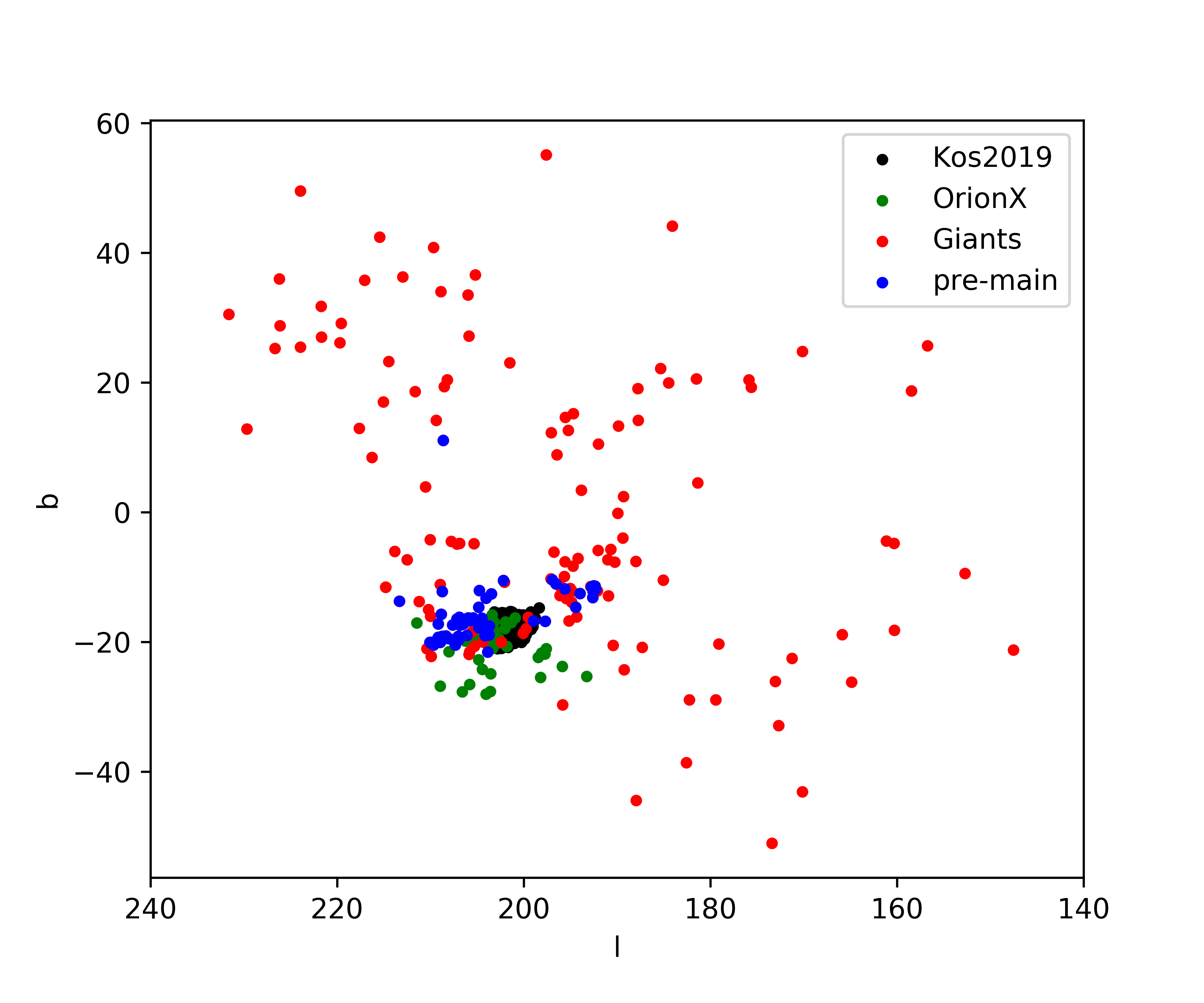}
\caption{Distribution of candidate members in galactic coordinate.\label{fig:lb}}
\end{figure}

\begin{figure}[tpb]
\plottwo{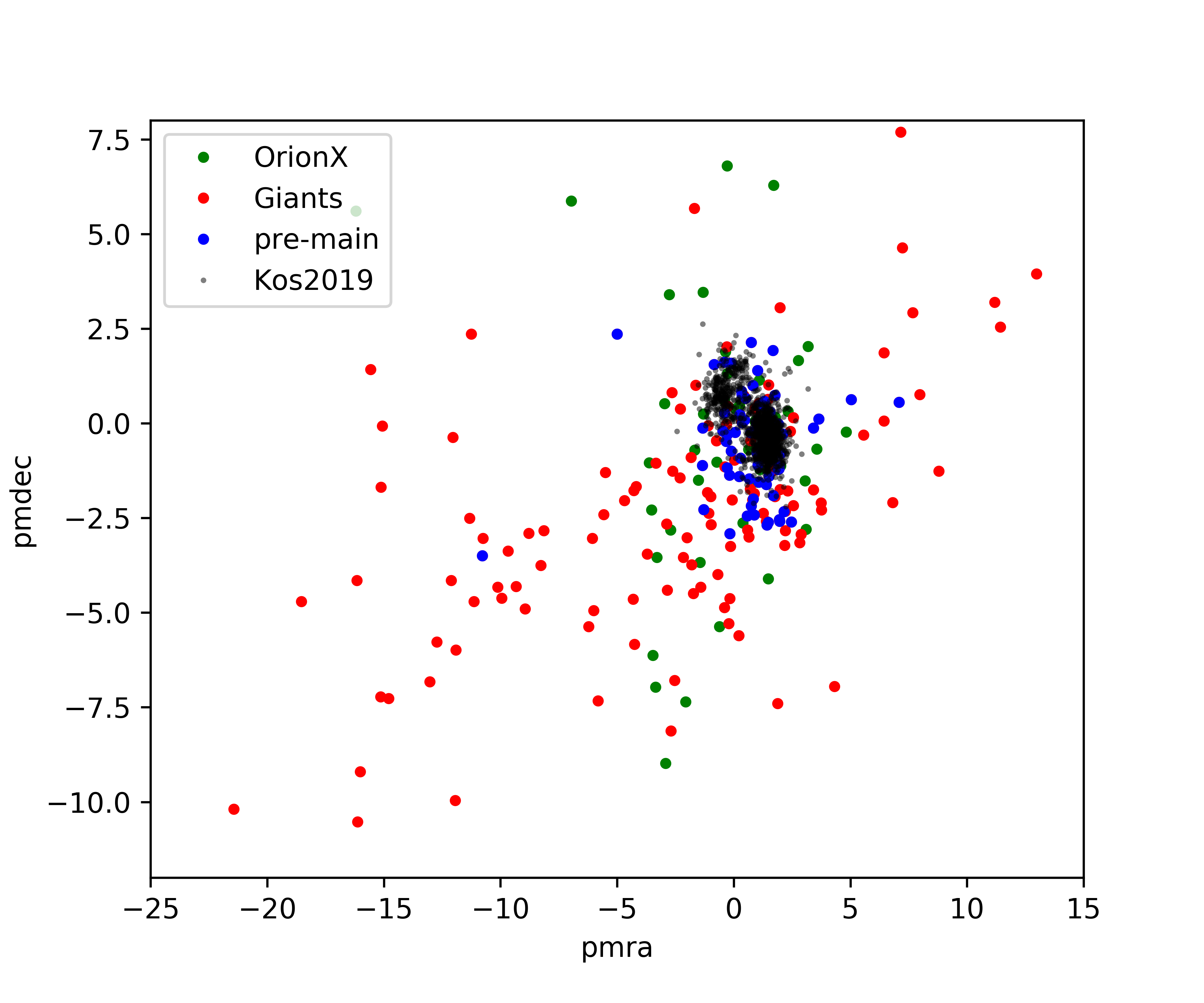}{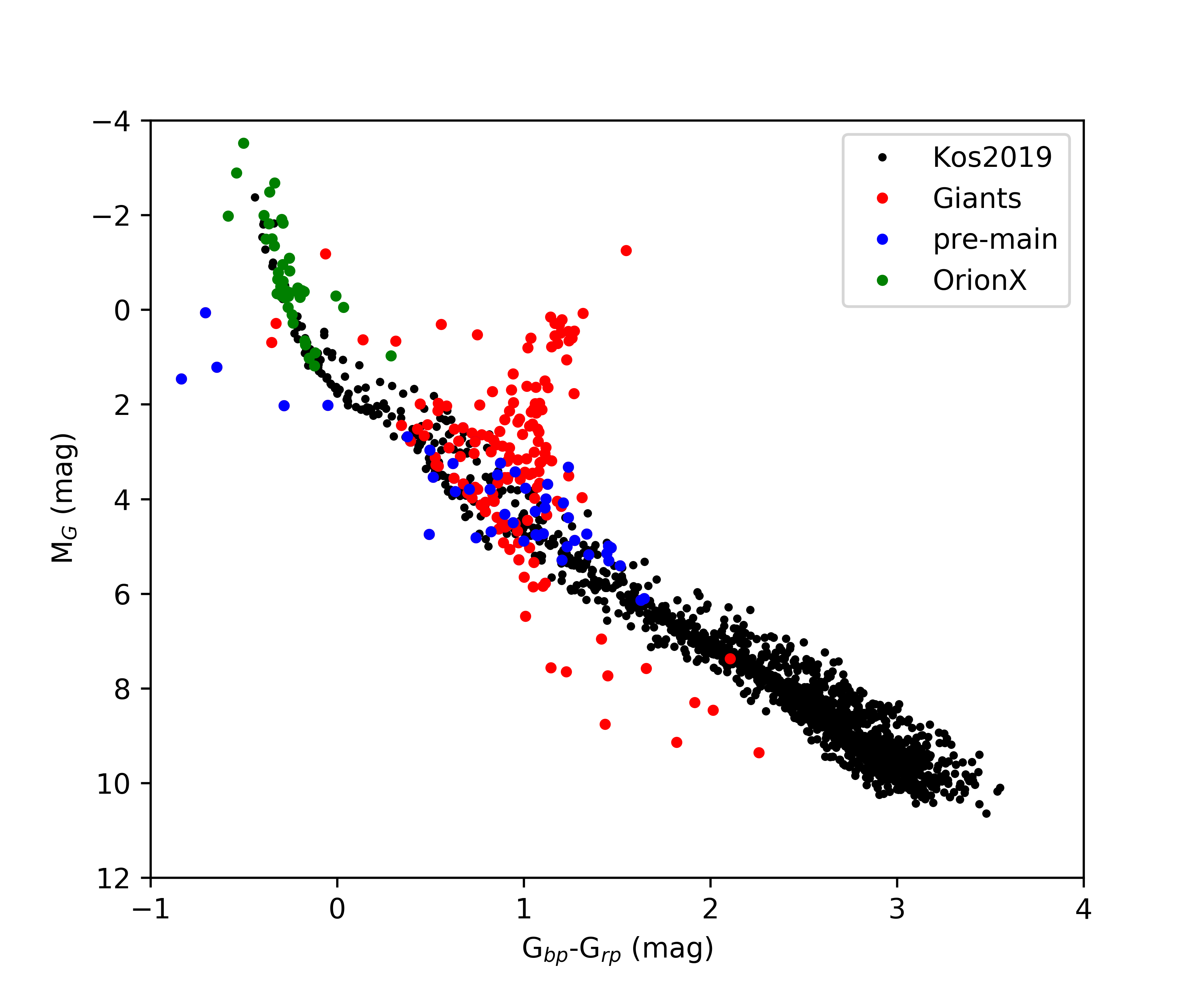}
\caption{Distribution of candidate members in proper motion coordinate and colour magnitude diagram.\label{fig:bprp}}
\end{figure}

Figure \ref{fig:lb} presents candidate members in galactic coordinate and the abscissa is galactic longitude and the ordinate is galactic latitude. Red dots represent normal giant stars and blue dots represent pre-main sequence stars while green dots represent member stars of OrionX from \citet{bou15}. OrionX is an young stellar moving group (mainly young O, B type stars) close to the Orion nebula. It is obvious that those pre-main sequence stars are very close to OrionX in the galactic coordinate and they are all inside the sky range of those G, K giants. It is known that moving groups can origin from diffusion/evaporation of clusters. Therefore we speculate that pre-main sequence stars of the new moving group are parts of star diffusion. Young stars move away from their born place along with old stars which gathered there by density wave, so that when the spiral arm move away, the old star born region can change into less dense as non-arm region. It is natural to think that when the peak of density wave comes, matter gathering together might have stimulated star forming events \textbf{ \citep{rob69}}.

\textbf{The left subplot of figure \ref{fig:bprp} presents candidate members in proper motion coordinate and the abscissa is proper motion in right ascension and the ordinate is proper motion in declination. Red dots represent normal giant stars while blue dots represent pre-main sequence stars of our candidate members. Green dots represent member stars of OrionX from \citet{bou15} while black dots represent member stars of ASCC 16, ASCC 18, ASCC 20, ASCC 21 and ASCC 21a from \citet{kos19}. There is one common pre-main sequence star between our candidate members and ASCC21. Stars from \citet{kos19} are basically enclosed by OrionX and our pre-main sequence stars which are enclosed by our giants. Our pre-main sequence stars have larger dispersion than \citet{kos19} because their stars were born in the orion nebula recently while our stars have started moving away. The right subplot of figure \ref{fig:bprp} presents candidate members in $M_G$ versus $G_{bp} - G_{rp}$ coordinate and $M_G$, $G_{bp}$ and $G_{rp}$ are absolute magnitudes. For stars from \citet{kos19}, we set extinction $A_v = 0.25$ as they did in that paper. For other stars, extinctions were obtained from Gaia dr2 catalogue if provided and if not provided, extinctions were obtained from Galactic Dust Reddening and Extinction website\footnote{http://irsa.ipac.caltech.edu/applications/DUST/} \citep{sch11}. The colour magnitude diagram shows that our pre-main sequence stars and OrionX may have similar age as stars from \citet{kos19}. Parts of our giants are very close to the distribution region of black dots while other are apparently field stars. In a word, we think pre-main sequence stars of the  new moving group came out of the orion nebula and they are moving away with OrionX and giants of the new moving group driven by density wave.}

\section{Conclusion}

In \textbf{summary}, we detected a new moving group from the \textbf{cross-matched} G, K giant star sample \textbf{between }LAMOST DR5 catalogue and Gaia DR2 catalog \textbf{and }candidate member stars gathered in spatial position, kinematics and chemical abundances coordinates. Their \textit{U, V} velocities distribute around a peak of wavelet transform coefficient of star count density distribution in that region. After the candidate members were picked out by convergent point method, they were divided into two stellar types by their positions in $T_{\textrm{eff}}$ versus log$g$ diagram. In those candidate members, young pre-main sequence member stars have been confirmed by spectra. According to its spatial position, we propose that this new moving group is possible to be related to the density wave of the Local Arm. We speculate that stars gathering together may trigger the first gravitational collapse in the star forming region. In the future, we hope to explore how density waves affect all those star forming regions.

\acknowledgments
 This study is supported by the National Natural Science Foundation of China under grant No. 11988101, 11973048, 11927804, 11890694, 11625313, 11803016 and National Key R\&D Program of China No. 2019YFA0405502. This work is also supported by the Astronomical Big Data Joint Research Center, co-founded by the National Astronomical Observatories, Chinese Academy of Sciences and the Alibaba Cloud. Guoshoujing Telescope (the Large Sky Area Multi-Object Fiber Spectroscopic Telescope LAMOST) is a National Major Scientific Project built by the Chinese Academy of Sciences. Funding for the project has been provided by the National Development and Reform Commission. LAMOST is operated and managed by the National Astronomical Observatories, Chinese Academy of Sciences.


\appendix

\begin{longdeluxetable}{cccc}
\tablecaption{Equivalent widths of lithium 6707 \AA \ line of 74 pre-main sequence stars of candidate members.\label{tab:liabu}}
\tablewidth{0pt}
\tablehead{\colhead{obsid} & \colhead{EW (m\AA)} & \colhead{snrr}}
\startdata
 46616064   &     310.88280 &  19.94 \\
 126816087  &     253.03027 &  16.21 \\
 160913149  &     689.66983 &  64.97 \\
 160909139  &     648.24384 &  284.38 \\
 181213112  &     473.50827 &  34.09 \\
 181206055  &     642.26395 &  38.91 \\
 181206106  &     549.58361  & 59.01 \\
 181206213  &     629.65890 &  60.85 \\
 181113058  &     727.05001 &  49.18 \\
 181113060  &     550.16684 &  156.96 \\
 181113089  &     1025.3223 &  245.74 \\
 181113178   &    758.18243 &  36.31 \\
 181113191  &     606.54541 &  70.90 \\
 181113198  &     566.03624 &  117.16 \\
 181212160  &     608.17286 &  69.52 \\
 181213007  &     576.49124 &  77.75 \\
 181213031  &     638.07816 &  112.63 \\
 181213047  &     632.88668 &  146.44 \\
 181213067  &     635.04579 &  31.71 \\
 181106007  &     623.39639 &  48.26 \\
 181106034  &     737.25087 &  24.74 \\
 181106208  &     447.25398 &  28.83 \\
 181109117  &     578.67973 &  89.76 \\
 207203063  &     693.99885 &  58.54 \\
 208802008  &     693.72804 &  113.81 \\
 257101241  &     702.36229 &  72.30 \\
 257105020  &     560.84465 &  65.57 \\
 257105025  &     561.82408 &  169.73 \\
 257107035  &     641.85997 &  125.44 \\
 257108111  &     518.76474 &  96.50 \\
 257108224  &     650.01448 &  61.42 \\
 259812150  &     635.79465 &  40.42 \\
 259806084  &     692.03167 &  125.15 \\
 259806108  &     669.38145 &  82.15 \\
 259809059  &     556.35304 &  105.06 \\
 259811032  &     615.41363 &  90.61 \\
 259803062  &     537.78091 &  166.00 \\
 268307043  &     559.58009 &  123.25 \\
 278015216  &     583.91663 &  56.74 \\
 278010250  &     618.95348 &  65.00 \\
 278006061  &     572.01251 &  52.67 \\
 278009145  &     614.77182 &  104.91 \\
 326912108  &     561.72492 &  25.50 \\
 326915140  &     579.15012 &  76.15 \\
 326915176  &     688.89391 &  116.99 \\
 326903204  &     693.84661 &  64.77 \\
 326904067  &     621.67493 &  122.00 \\
 326904112  &     508.80706 &  128.95 \\
 326905016  &     626.24413 &  58.98 \\
 326905022  &     726.81521 &  241.10 \\
 326905144  &     638.50620 &  61.58 \\
 326908012  &     755.03301 &  98.52 \\
 326908214  &     626.96787 &  57.48 \\
 326908236  &     617.08667 &  113.77 \\
 326901138  &     645.91023 &  82.99 \\
 326901146  &     766.15445 &  52.64 \\
 326903103  &     610.89172 &  82.91 \\
 326903158  &     586.26447 &  74.13 \\
 427101018  &     529.39069 &  50.73 \\
 483615045  &     660.70939 &  196.68 \\
 487715184  &     602.20676 &  177.54 \\
 495514094  &     588.62183 &  47.00 \\
 495516045  &     619.31001 &  57.44 \\
 495503192  &     656.06453 &  69.57 \\
 505210149  &     660.85759 &  234.05 \\
 505215016  &     559.57665 &  223.75 \\
 505215049  &     623.60200 &  216.21 \\
 505206040  &     629.24914 &  214.47 \\
 505209051  &     658.80352 &  79.08 \\
 505209081  &     682.32856 &  98.52 \\
 505209122  &     713.66930 &  134.43 \\
 553606212  &     599.17672 &  119.65 \\
 553608180  &     699.95209 &  150.53 \\
 553608236  &     642.95106 &  52.61 \\
\enddata
\end{longdeluxetable}

\newpage
\begin{figure}
\plotone{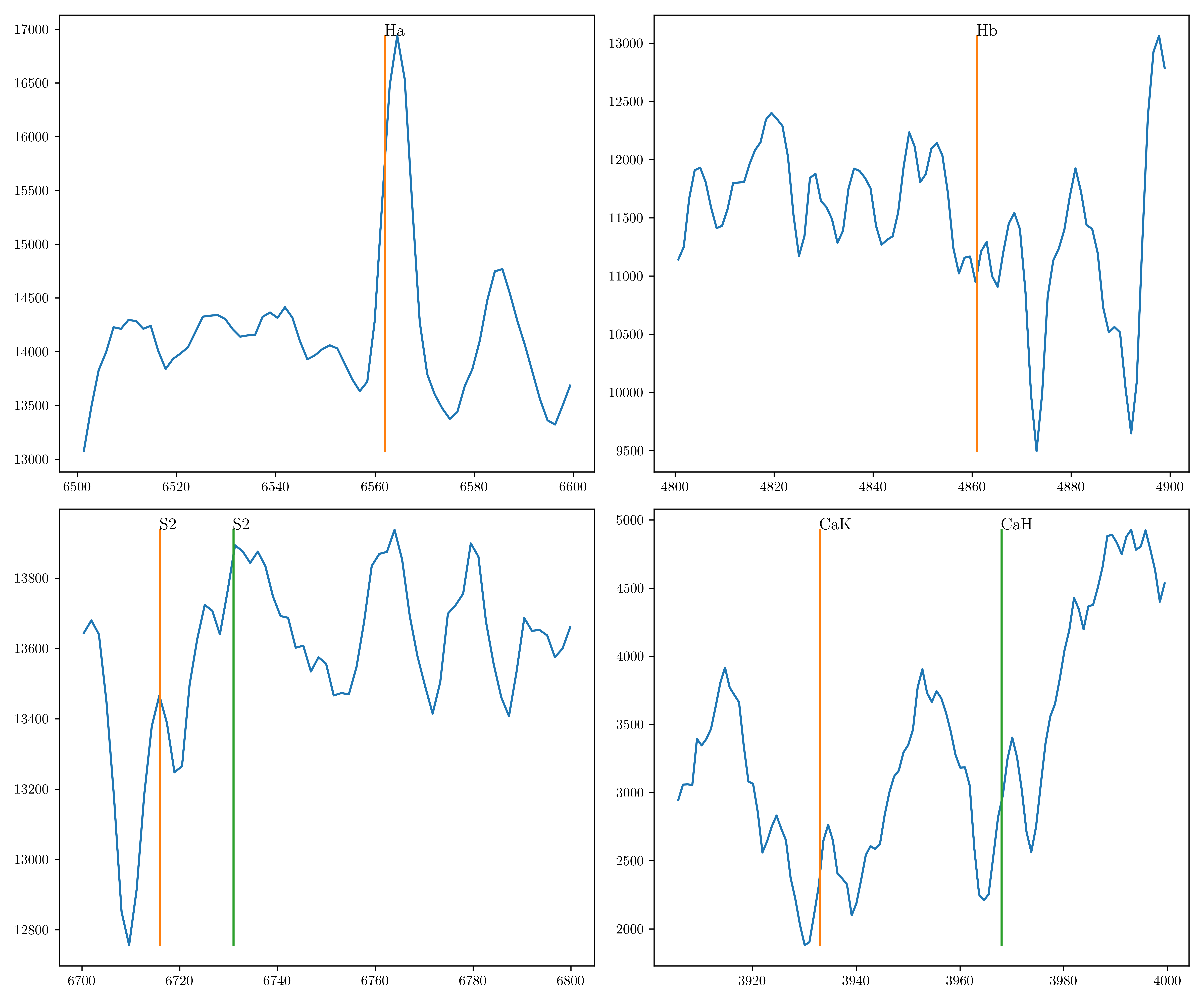}
\caption{A example of LAMOST low resolution spectrum features of a candidate pre-main sequence star of new moving grou of p.\label{fig:gp}}
\end{figure}

\end{document}